\documentclass[11pt]{article}

\usepackage{graphicx}
\usepackage[squaren]{SIunits}
\usepackage{isolatin1}
\usepackage{hyperref}

\newcommand{\dd}{\mbox{d}}
\newcommand{\eff}{\mbox{eff}}
\newcommand{\QED}{\mbox{QED}}

\begin{document} 

\title{HARPO, a gas TPC active target for high-performance $\gamma$-ray astronomy; demonstration of the polarimetry of MeV $\gamma$-rays converting to $e^+ e^-$ pair}

\author{D.~Bernard, on behalf of the HARPO Collaboration.
\\
LLR, Ecole Polytechnique, CNRS/IN2P3, 91128 Palaiseau, France}

\maketitle 

\begin{center}
\large \textbf{Presented at PM2018,
\\
 14th Pisa Meeting on advanced detectors, 27 May - 02 June 2018, Isola d'Elba, Italy }
\end{center}

\begin{abstract}
No $\gamma$-ray polarimeter sensitive above 1 MeV has ever flown space.
$\gamma$-ray polarimetry would be a new window on the radiative processes
at work in cosmic sources, processes that produce linearly polarised
emission, each of which with different polarisation fractions.
The HARPO Collaboration has designed, built and characterised on beam
a gas-TPC active target with which we have demonstrated for the first
time the polarimetry of a linearly polarised MeV $\gamma$-ray beam, from
the analysis of the conversions to $e^+ e^-$ pairs.
\end{abstract}

{\em keywords }:
$\gamma$-rays, polarimeter, TPC, pair production, telescope, gas detector

\section{$\gamma$-ray polarimetry of cosmic sources: a science case}

Today's $\gamma$-ray astronomy is polarisation-blind as no significant
polarimetry of a cosmic source has been performed.
Polarimetry would provide the additional observables needed to solve the
pulsar (angle configuration, magnetic field, emission location)
parameter degeneracy.
In particular in MeV pulsars, it would enable to
decipher the high-energy $\gamma$-ray emission mechanism
(curvature radiation or 
synchrotron radiation) \cite{Harding:2017tdm,Harding:2017ypk}.
Polarimetry 
 would enable to identify the nature of the emitting particles in $\gamma$-ray
emitting blazar jets (pure leptons or lepton-hadron mixture)
\cite{Zhang:2013bna}.
Polarimetry could enable the indirect detection of dark
matter \cite{Huang:2018qui};
the detection of Lorentz-invariance violation due to ``new'' physics
beyond the standard model, with a potential larger than that of past searches in
the X-ray band as the sensitivity varies with the square of the
photon energy \cite{Stecker:2011ps};
the discovery of the axion, i.e. of the pseudo-scalar believed to be
the pseudo-Goldstone boson associated to a breaking of the $U(1)$
symmetry devised to solve the QCD CP problem \cite{Rubbia:2007hf}.

\section{$\gamma$-ray polarimetry of cosmic sources: how achieve it ? }

For the lowest part of the electromagnetic spectrum, polarimetry is
performed by the measurement and the analysis of light intensities or
of electric fields.
At high energies, the (photo-electric, Compton scattering or
conversion to pair) interaction of each single photon in a detector is
recorded, and an estimate of the azimutal angle, $\phi$, of the particles in
the final-state is measured.
The expression of the cross section, differential with respect to $\phi$, is 
\begin{equation}
 \dd \sigma / \dd \phi \propto \left(1 + A \times P \cos(2(\phi-\phi_0)) \right)
 ,
\end{equation}
where 
\begin{itemize}
\item $A$ is the polarisation asymmetry of the process at work. Its value depends on the photon energy.
\item $P$ is the linear polarisation fraction of the radiation.
\item $\phi_0$ is the angle between the polarisation direction of the radiation and the reference axis with respect to which the azimutal angle is measured.
\end{itemize}
The product $ A \times P $ is sometimes named the modulation factor.

The polarimetry of a cosmic source in the MeV-GeV energy range
has never been performed.
For Compton polarimeters, both the total cross-section and the polarisation
asymmetry decrease with energy 
(\href{http://inspirehep.net/record/1382323/files/acompton3.png}{Fig. 2}
of \cite{Bernard:2015hwa}).
For nuclear pair-conversion, $\gamma Z \to e^+ e^- Z$,
the multiple scattering undergone by the
electron and by the positron in the tracker blur the azimuthal-angle
information carried by the pair within $\approx 10^{-3}$ radiation
lengths \cite{Kelner,Kotov,Mattox}.

Hopes have been put in the use of triplet conversions, that is, of
conversions to pair in the field of an electron of the detector,
$\gamma e^- \to e^+ e^- e^-$, as the target electron oftens recoils at
a large polar angle with respect to the direction of the incoming
photon, making it easier the measurement of its azimuthal angle.
Alas, the cross section is small and the fraction of it for
high-enough recoil electron momentum is even much smaller
(\href{http://inspirehep.net/record/1242601/files/depaola-bis.png}{Fig. 6}
of \cite{Bernard:2013jea}), so the potential for a measurement
on a cosmic source ends up to be miserable (Sect. 5.3
of \cite{Bernard:2013jea}).

A way to mitigate the multiple scattering issue is the use of
high-imaging-resolution active targets such as emulsion detectors
\cite{Takahashi:2015jza}.
Zooming in the first microns of the event just downstream
 the conversion vertex,
polarimetry with conversions to pairs was demonstrated with a
prototype detector in a GeV $\gamma$-ray beam \cite{Ozaki:2016gvw} but
the ability to take data in the MeV energy range, where most of the
statistics lies for cosmic sources (Fig. 2 of \cite{Bernard:2013jea}),
remains an issue.

The HARPO project explored an other technique, the use of a
low-density, homogeneous active target such as a high-pressure gas TPC
(time projection chamber) \cite{Bernard:2014kwa}.
We first wrote \cite{Bernard:2013jea} a Monte-Carlo event
generator\footnote{A C++ version
of our event generator \cite{Igor:2018} has been deployed as the
G4BetheHeitler5DModel physics model of the 10.5beta release of Geant4
\cite{Agostinelli:2002hh,Allison:2016lfl}
(A fortran demonstrator is documented in \cite{Bernard:2018hwf}).}
that samples the five-dimensional (5D) differential cross
section at the first order of the Born development, i.e., the
``Bethe-Heitler'' differential cross section (non polarised
\cite{Bethe-Heitler}, polarised
\cite{BerlinMadansky1950,May1951,jau}).
With that tool we determined the (68\,\% containment) contribution to
the single-photon angular resolution due to the fact that the momentum
of the recoil nucleus cannot be measured, to be
 $\approx 1.5 \, \radian [E / 1\, \mega\electronvolt]^{-5/4}$ 
\cite{Bernard:2012uf,Gros:2016zst}.
We obtained the contribution to the single-photon angular resolution
due to the single-track angular resolution in the case of an optimal
tracking \`a la Kalman to be $\propto p^{-3/4}$
\cite{Bernard:2012uf,Frosini:2017ftq}.

At high thicknesses, we found the dilution\footnote{The dilution
 factor $D$ is the factor by which the polarisation asymmetry in an
 actual detector, $A_{\eff}$, is decreased with respect to the
 theoretical value, $A_{\QED}$, that is, $A_{\eff} = D \times
 A_{\QED}$.}
of the polarisation asymmetry due to multiple scattering to worsen
less (Fig. 17 of \cite{Bernard:2013jea}), with this full Monte-Carlo
event generation, than what was predicted by
\cite{Kelner,Kotov,Mattox} who approximated the pair opening-angle,
$\theta_{+-}$, by its most-probable value.
This is due to the long tail of the $\theta_{+-}$ distribution.
We demonstrated that the single-track angular resolution of gas
detectors can be so good that polarimetry can be performed before the
azimuthal information is lost \cite{Bernard:2013jea}.
Due to the $p^{-3/4}$ variation of the single-track angular
resolution,
and given the $1/E$ scaling of the $\theta_{+-}$ distribution
\cite{Olsen1963},
a better dilution is within reach at lower energies
(Fig. 20 of \cite{Bernard:2013jea}).
The simulation shows that with a $1\,\meter^3$, 5\,bar argon polarimeter
observing a cosmic source such as the Crab nebula for one year, full
time, with full efficiency, the expected precision in the measurement
of $P$ is of $\approx 1.4\,\%$ \cite{Bernard:2013jea}.

\begin{figure}[htb]
 \centerline{
 \includegraphics[width=0.62\linewidth]{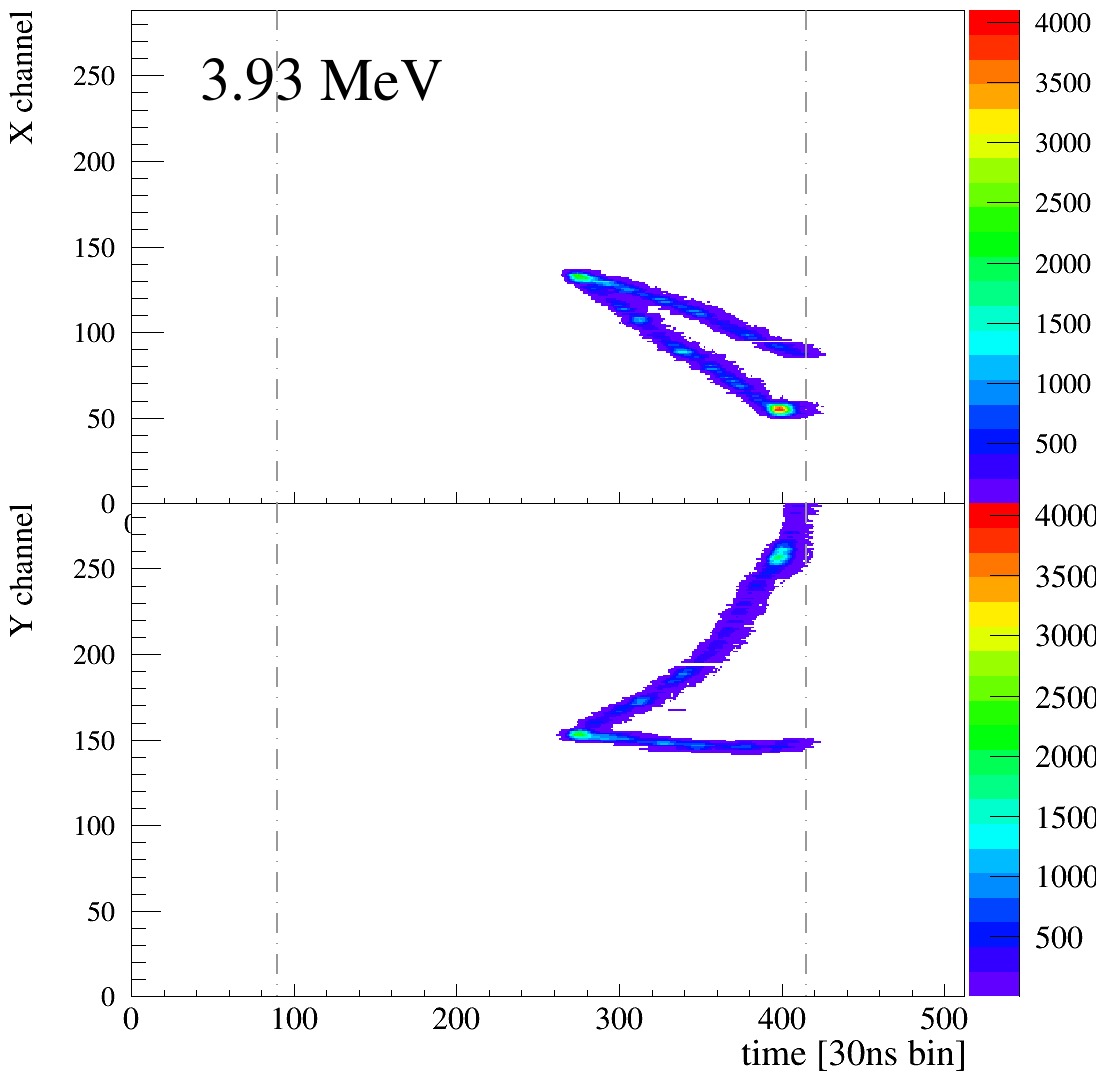}
 }
 \caption{The two ``maps'', that is, the two $x,t$ and $y,t$ projections of a conversion event of a 3.93\,MeV $\gamma$-ray provided by the NewSUBARU BL01 beam line converting to an $e^+e^-$ pair in the 2.1\,\bbar\ argon-isobutane (95-5\,\%) gas of the HARPO TPC prototype \cite{Bernard:2017jmw}.
 The vertical dashed lines denote the physical limits of the detector, i.e. the cathode (left) and the anote (right).
\label{fig:a}
 }
\end{figure}

We implemented the moments' method to obtain an optimal measurement of
the polarisation modulation factor $A \times P$.
In the case of the simple 1D differential cross section, the method is
equivalent to a maximum likelihood fit of the 1D differential cross section
$\dd \sigma / \dd \phi \propto \left(1 + A \times P \cos(2(\phi-\phi_0)) \right)$
\cite{Bernard:2013jea,Gros:2016dmp}.
In the case of the full 5D differential cross section, an additional
gain in the precision of the measurement of $A \times P$ of a factor of two-to-three is obtained
\cite{Bernard:2013jea,Gros:2016dmp}.

In the case of interest of a final state consisting of three particles
(even when the recoil cannot be measured), the definition of the
azimuthal angle, $\phi$, is an issue.
We determined the bisector $\phi_{+-} = (\phi_{+} + \phi_{-})/2$ of the
azimuthal angles of the direction of the electron, $\phi_{-}$, and of
the positron, $\phi_{+}$, to be the optimal choice
\cite{Gros:2016dmp}.
We demonstrated that with that angle definition, our generator is the
only one, to our knowledge, to match the known
asymptotic expressions for $A$ at low and at high
energies \cite{Gros:2016zst}.

\section{HARPO: the experiment }

The HARPO Collaboration built a high-pressure (0.5 - 5 bar) gas TPC
prototype
\cite{Bernard:2014kwa} that uses an hybrid GEM-micromegas amplification
system that we characterised precisely \cite{Gros:TIPP:2014}.
The anode plane is segmented into 2 orthogonal series ($x,y$) of strips,
rather than into pads, so as to limit the number of electronic channels and
therefore the power consumption, a key factor for a detector intended
to fly a space mission.
The ambiguity in the track assignment in the two ($x, t$ and $y, t$)
event projections, where $t$ is the ionisation electron drift time, is easily solved by
matching the charge time-distribution for each track (Fig. 6 of
\cite{Bernard:2012jy}).

\begin{figure}[htb]
 \centerline{
 \includegraphics[width=0.8\linewidth]{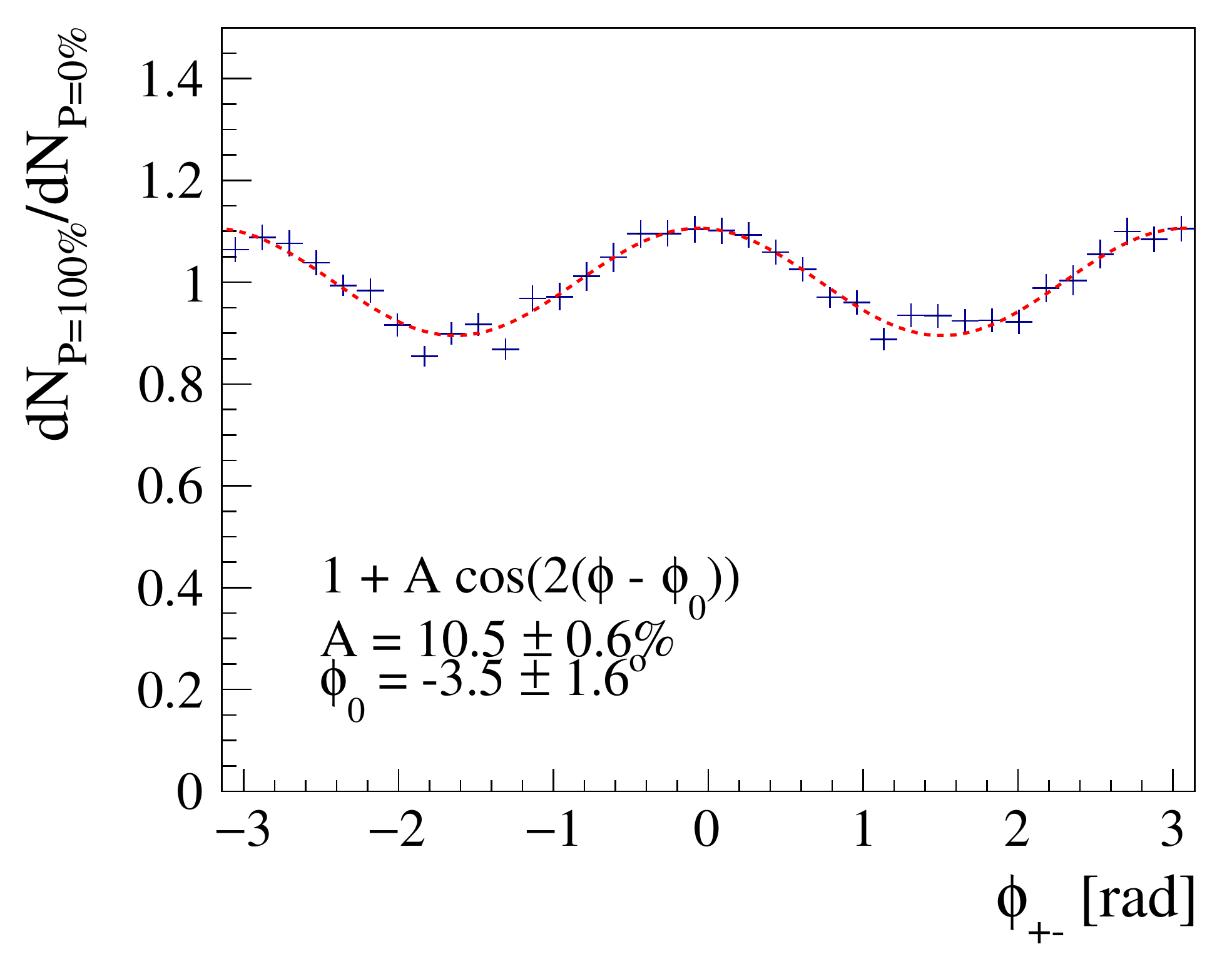}
 }
 \caption{Distribution of the azimuthal angle of 11.8\,MeV $\gamma$-rays (ratio of the fully linearly polarised to the linearly non-polarised) converting to an $e^+e^-$ pair in the 2.1\,\bbar\ argon-isobutane (95-5\,\%) gas of the HARPO TPC prototype \cite{Gros:2017wyj}.
\label{fig:b}
 } 
\end{figure}

The detector was exposed to a high-intensity $\gamma$-ray beam
produced by the head-on inverse Compton scattering of a laser beam on the
0.6 - 1.5\,GeV electron beam of the NewSUBARU storage ring
\cite{Delbart:2015rmp}.
The $\gamma$-ray beam was collimated on axis so as to select forward
scattering and to obtain a quasi-monochromatic beam with photon
energies close to the Compton edge.
After collimation, the linear polarisation of the laser beam is
transferred almost perfectly to the $\gamma$-ray beam.
By varying the laser wavelength and the electron beam energy we were
able to take data between 1.7 and 74 MeV.
We used a dedicated trigger system that enabled the data taking of
$\gamma$-ray conversions in the TPC gas at rates of several 10\,Hz, in
the presence of several 10\,kHz of background noise, with a negligible
dead time \cite{Geerebaert:2016dyv}.
We simulated the experiment with a Geant4-based Monte-Carlo
simulation, with a custom implementation of the TPC-specific processes
that we calibrated with respect to beam data \cite{Gros:TPC:2016}.

The strongly non-cylindrical-symmetric $x,y,t$ structure of the detector
 induces a bias in the $\phi$ distribution:
 the distribution angle of a non-polarised sources is not flat 
(Fig. 9 of \cite{Gros:2017wyj}).
This was analysed and mitigated by a two-step strategy:
(a) chunks of data were taken with the detector rotated around the
beam axis by multiples of $45^\circ$;
(b) data were taken with either fully polarised ($P=1$) or non
polarised ($P=0$) beam, so as to form the ratio of the $\phi$
distributions;
On a space mission, a non-polarised beam would not be available, and
therefore $P=0$ simulated samples would be used to apply the bias
correction.
We demonstrated that this can be done without any additional
systematic bias due to a possible difference between simulation and
data, within uncertainty (Fig. 11 of \cite{Gros:2017wyj}).
The biases mentioned above would average out for an isotropic exposure
anyway, as is the case for a long-duration data taking, so the
MC-based correction considered above would be a second-order
correction (Fig. 12 of \cite{Gros:2017wyj}).

The dilution factor is found to be of about 0.4 at 11.8\,MeV and to
decrease with photon energy, as expected
(Fig. 13 of \cite{Gros:2017wyj}).

\section{Results}

We have demonstrated for the first time the polarimetry of a MeV
$\gamma$-ray beam and with an excellent dilution factor
\cite{Gros:2017wyj} (Fig. \ref{fig:b}).
In contrast to higher-density targets, gas detectors are able to
image the conversion to pairs of low-energy
$\gamma$-rays (Fig. \ref{fig:a}).
This is of utmost importance as the polarisation asymmetry of
conversions to pair does not decrease much on the energy range of a
practical polarimeter\cite{Bernard:2013jea,Gros:2016dmp}:
the precision of the measurement relies ultimately on the
size of the collected statistics and therefore on the ability to
collect data efficiently at low energies.

An attempt to monitor the degradation of the TPC gas in a sealed mode
over half a year has shown that after the oxygen contamination (with
an oxygen/nitrogen fraction found to be compatible with air) was
removed by circulation through a purification cartridge, the nominal
TPC fresh-gas parameters are restored \cite{Frotin:2015mir}, enabling
a long-duration operation with the same gas on a space mission.

My gratitude to the French National Research Agency for her support 
(ANR-13-BS05-0002).

\bibliographystyle{elsarticle-num}

\end{document}